\documentclass[prd,aps,reprint,superscriptaddress,nofootinbib]{revtex4-2}

\pdfoutput=1

\usepackage[colorlinks=true,urlcolor=blue,anchorcolor=blue,citecolor=blue,filecolor=blue,linkcolor=blue,menucolor=blue,linktocpage=true,pdfproducer=medialab,pdfa=true]{hyperref}
\usepackage[T1]{fontenc}
\usepackage{graphicx}
\usepackage{enumitem}
\usepackage{latexsym}
\usepackage{amsfonts}
\usepackage{amssymb}
\usepackage{mathrsfs}
\usepackage{color}
\usepackage{amsmath}
\usepackage[capitalise]{cleveref}
\usepackage{slashed}
\usepackage{dcolumn}
\usepackage{verbatim}
\usepackage{comment}
\usepackage{float}
\usepackage{multirow}
\usepackage{xspace}
\usepackage[dvipsnames]{xcolor}
\usepackage[normalem]{ulem}

\newcommand{\keV}{~\text{keV}}
\newcommand{\MeV}{~\text{MeV}}
\newcommand{\GeV}{~\text{GeV}}

\def\pitt{PITT PACC, Department of Physics and Astronomy,\\ University of Pittsburgh, 3941 O’Hara St., Pittsburgh, PA 15260, USA}

\begin{document}

%%%%%%%%%%%%%%%%%%%
\title{Exothermic Dark Matter at LZ}
%%%%%%%%%%%%%%%%%%%

\author{Carlos Henrique de Lima}
\email{carlosdelima@pitt.edu}
\affiliation{\pitt}

\begin{abstract}
We investigate whether the high-energy nuclear recoil reported by the LUX-ZEPLIN (LZ) experiment can be explained by exothermic dark matter. We realize this framework in a minimal inelastic dark photon model and show that both freeze-out and low-reheating-temperature freeze-in cosmologies can reproduce the observed rate, with GeV-scale mediator masses and small kinetic mixing. We find that the LZ event can occur in a preferred parameter region with dark matter mass around $m_\chi \sim 30-200~\mathrm{GeV}$, and mass splitting $\delta \sim 0.5-1~\mathrm{MeV}$. The scenario is largely insensitive to halo uncertainties and has robust predictions for different targets. In particular, argon and germanium detectors are expected to observe events at energies above the current xenon window.
\end{abstract}

\maketitle

%%%%%%%%%%%%%%%%%%%%%%%%%%%%%%%%%%%%%%%%%%%%
\section{Introduction} 
\label{sec:intro}
%%%%%%%%%%%%%%%%%%%%%%%%%%%%%%%%%%%%%%%%%%%%

It is currently understood that there is an important fraction of our universe that either does not interact with the Standard Model or is at least weakly interacting. So far, dark matter evidence has been observed only gravitationally, making the microscopic properties of dark matter difficult to pin down. This may be on the verge of changing, with the recent result from the LUX-ZEPLIN (LZ)  experiment, reporting one event (not currently explained under any of their backgrounds) with nuclear recoil $E_R=248 \pm 23 \pm 23$ keV in a 2.84 tonne-year exposure and global significance of 2.6$\sigma$~\cite{LZ:2026axp}. One event is too few, and with significance far below the gold standard of $5\sigma$, it is too early to conclude. Nevertheless, it is interesting to ask what constructions could generate this signature and are consistent with other experimental searches.

The local behavior of the event, without any lower-energy recoil, makes ordinary elastic scattering interpretations unlikely. The hard scattering event would also be accompanied by several events at lower energies, where the detector would have enough sensitivity to measure them. This gives a hint that, if this event is indeed dark matter, it may be coming from an inelastic process~\footnote{Realizations of elastic effective field theory operators were explored in Ref.~\cite{LZ:2026axp} with UV completion in terms of axion portal in Ref.~\cite{Unwin:2026rdp}.}. 

Considering inelastic dark matter~\cite{Tucker-Smith:2001myb}, there could be two ways that would describe this signature: endothermic or exothermic dark matter. Endothermic dark matter, where the event is an up-scattering, was covered by the LZ effective-field-theory scan and explored in the higgsino interpretations that followed~\cite{DiMauro:2026ldr,Lou:2026idn,Wu:2026nhi,Freese:2026sga,Fan:2026kxx,Du:2026guj}~\footnote{Other realizations were proposed as from a Peccei-Quinn mechanism~\cite{Visinelli:2026kgt,Yin:2026jnn}, neutrino up-scattering~\cite{Jeesun:2026vzo},  dark photon dark matter~\cite{Yamashita:2026ump} and $Z_2$ Higgs doublet~\cite{Nomura:2026qyq}. }. This mechanism can reproduce the recoil event, and in the case of Higgsino dark matter, does so naturally with not so much freedom in the parameters. The main challenge of this description is that, to deposit around 250 keV of energy with a splitting of a few hundred keV, the velocity of this dark matter event lies in the tail of the halo distribution, where there is a lot of uncertainty~\cite{Smith-Orlik:2023kyl}. It may also be in tension with neutrino emission from solar dark matter capture~\cite{Pospelov:2026ewn} from IceCube measurements. 

In this work, we propose the other option. Consider that a fraction of the halo lives in a long-lived metastable state $\chi_H$~\cite{Finkbeiner:2007kk}. The down-scattering transition $\chi_H N \to \chi_L N$ then releases the mass difference $\delta$ into the collision, and this population does not need to sit in the tail of the halo velocity distribution to generate a high recoil event. The excited state must still be present today, implying $\delta < 2m_e$~\footnote{The exothermic construction was also considered independently in the concurrent work of Ref.~\cite{Dent:2026bji}, which compares it with the endothermic explanation and invokes a leptophobic mediator to allow larger splittings.}. Down-scattering with keV to MeV splittings has been studied before \cite{Graham:2010ca,Batell:2009vb,Finkbeiner:2009mi,CarrilloGonzalez:2021lxm}, and pseudo-Dirac dark sectors of this kind have been built and searched for in several places \cite{Cui:2009xq,Izaguirre:2015zva,Finkbeiner:2009mi,Blennow:2016gde,Bhattiprolu:2023akk,Brahma:2023psr}. We map out the preferred parameter space from the LZ measurement and compare it to current and near-future experiments, considering also the dark matter cosmological history. We find that the window of dark matter mass $m_\chi \sim 30$–$200~\mathrm{GeV}$, and mass splitting $\delta \sim 0.5$–$1~\mathrm{MeV}$ can explain the LZ measurement, for GeV-scale mediators with kinetic mixing of the order of $10^{-6}$.

%%%%%%%%%%%%%%%%%%%%%%%%%%%%%%%%%%%%%%%%
\begin{figure*}[t!]
\centering
\includegraphics[width=2\columnwidth]{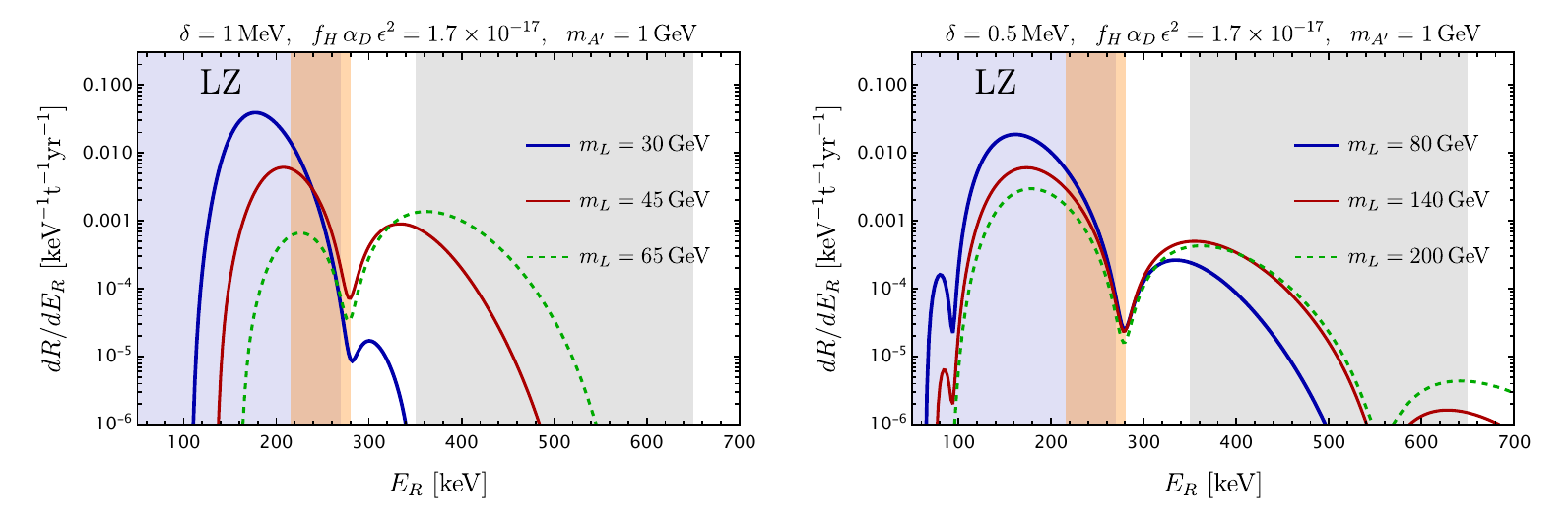}
\caption{Recoil spectra in xenon from Eq.\eqref{eq:rate}. The splitting is held fixed in each panel, at $1.0\MeV$ on the left and $0.5\MeV$ on the right, and the mass is varied over the allowed range. All six curves share the same $f_H\alpha_D\epsilon^2 = 1.7\times 10^{-17}$ and $m_{A'} = 1$ GeV. Light shading is the LZ window, $5.4-270\keV$, dark shading is the sideband, $350-650\keV$, and the orange band is the reported energy of the event.}
\label{fig:rates}
\end{figure*}
%%%%%%%%%%%%%%%%%%%%%%%%%%%%%%%%%%%

%%%%%%%%%%%%%%%%%%%%%%%%%%%%%%%%%%%
\section{Exothermic scattering}
%%%%%%%%%%%%%%%%%%%%%%%%%%%%%%%%%%%

Assuming a fractional population of excited dark matter $f_H$, composed of a two-state system with splitting $\delta = m_H-m_L$, the minimal velocity as a function of the nuclear recoil for the down-scattering  $\chi_H N \rightarrow \chi_L N$ process is 
\begin{align}
    v_{\rm{min}}(E_R) = \frac{1}{\sqrt{2m_N E_R}} \left| \frac{m_N E_R}{\mu} -\delta\right|\, ,
\end{align}
where $m_N$ is the target nucleus mass and $\mu$ the reduced dark matter nucleus mass. The minimal velocity vanishes for $E_0=\frac{m_\chi \delta}{m_\chi + m_N}$, meaning that the differential rate is peaked around $E_0$~\cite{Lang:2010cd}. The accessible recoil energies at speed $v$ are
\begin{align}
E_\pm(v) = \frac{\mu^2 v^2}{2 m_N}\left[ 1 \pm \sqrt{1 + \frac{2\delta}{\mu v^2}} \right]^2 \, .
\label{eq:Epm}
\end{align}
The differential distribution is then expected to be centered at $E_0$ with fractional half-width $w =2 v \sqrt{\frac{\mu}{2\delta}}$. The LZ extension of the region of interest from 55 keV to 270 keV is the first exposure to a large fraction of this allowed window. Requiring $E_0$ close to the observed energy restricts the lower end of the dark matter mass to $m_\chi \simeq 30\GeV$~\footnote{Since $\delta \ll m_L$ we write the dark matter mass as $m_L$ throughout, with $m_H = m_L + \delta$, and $m_\chi=m_L$.}.

The splitting $\delta$ is directly connected to the lifetime of the heavy state $\chi_H$~\cite{Batell:2009vb,Krnjaic:2025zjl}. Generically, the same off-diagonal vertex that generates the nuclear recoil can also decay the excited population away: $\chi_H \rightarrow \chi_L f \bar{f}$ with an off-shell dark mediator. Consider a dark photon with coupling $g_D$ and $\epsilon e$ to dark matter and SM fermions, respectively. The three-body decay forces the splitting to be smaller than the electron-positron pair production, $\delta < 2m_e$~\footnote{Additional model building can be done to push this further above with species-specific mediators.}. Even when this channel is closed, there are still two possible decay modes, where it goes to neutrinos or through photons. These are suppressed by additional powers of small couplings and are slower than the Hubble rate for the values of couplings explored in this work. Let us now focus more on the inelastic dark photon scenario.

%%%%%%%%%%%%%%%%%%%%%%%%%%%%
\section{Inelastic Dark Photon}
%%%%%%%%%%%%%%%%%%%%%%%%%%%%%

Consider a Dirac fermion $\chi$ charged under a dark abelian gauge symmetry with gauge boson $A^{'}$, with a kinetic mixing $\epsilon$. A dark Higgs or other mechanism generates a Majorana mass for $\chi$, splitting the eigenstates into $\chi_H$ and $\chi_L$. The gauge interaction becomes purely off-diagonal,
\begin{align}
    \mathcal{L} \supset i g_D A_\mu^{'} \bar{\chi}_L \gamma^{\mu} \chi_H \, ,
\end{align}
and the dark photon couples to the Standard Model with strength $\epsilon e$. This simplified model has five free parameters $\{\, m_L,\ \delta,\ m_{A'},\ g_D,\ \epsilon \,\}$, where some are fixed by the dark matter cosmological history and others are constrained by the observed event in the LZ recoil window.

Each vertex flips $\chi_L \leftrightarrow \chi_H$, so a single dark photon exchange is always inelastic. Elastic scattering needs two dark vertices, and it first appears at second order in the portal through a virtual $\chi_H$~\cite{Batell:2009vb}. The differential cross section for down-scattering on a nucleus of charge $Z$ is
\begin{align}
    \frac{d\sigma}{d E_R} = g_D^2\epsilon^2e^2Z^2\frac{m_N}{2\pi v^2 (2m_NE_R+m_{A^{'}}^2)^2} F^2(2m_NE_R) \, ,
\end{align}
and for a halo population $\rho_H = \rho_\chi f_H$ the recoil rate per unit detector mass is
\begin{align}\label{eq:rate}
    \frac{dR}{dE_R} =g_D^2\epsilon^2e^2Z^2 \frac{\rho_H}{m_\chi} \frac{F^2(2m_NE_R)}{2\pi (2m_NE_R+m_{A^{'}}^2)^2} \int_{v>v_{\rm min}} d^3v \frac{f(\boldsymbol{v})}{v} \, ,
\end{align}
where the integral is the usual halo velocity integral. Note that the rate is sensitive to the combination $\alpha_D \epsilon^2 f_H$, where $\alpha_D = g_D^2/4\pi$.

The rate is controlled by the nuclear form factor and the halo velocity integral. The halo integral is flat-topped at $E_0$,  while the coherent xenon response has its second diffraction minimum at $E_R \simeq 280\keV$, and falls by two orders of magnitude between $200$ and $270\keV$. For the halo integral, we use  Ref.~\cite{Baxter:2021pqo}, $\rho_\chi = 0.3\GeV\,{\rm cm^{-3}}$, a Maxwellian with $v_0 = 238$ km/s truncated at $v_{\rm esc} = 544$ km/s, and an Earth speed of $254$ km/s. $F^2$ is the Helm form factor in the Lewin-Smith parametrization \cite{Lewin:1995rx}, summed over the natural isotopes of each target. The rate distributions for different choices of parameters can be seen in Fig.~\ref{fig:rates}.

To select the best values of parameters that explain the LZ measurement, we construct an extended unbinned likelihood for exactly one event at the observed energy inside the LZ window and zero events everywhere else, including the sideband window of $350-650$ keV, which we obtain by converting the LZ high-energy sideband in $S1c$ into recoil energy. Ref.~\cite{Rodd:2026tyn} uses the same absence of events there to constrain the higgsino interpretation. We use the $2.84$ tonne-year exposure and the $5.4-270\keV$ window of Ref.~\cite{LZ:2026axp}. The event sits at $E_R = 248\keV$, and we combine its two quoted uncertainties into $\sigma = 32.5\keV$.

%%%%%%%%%%%%%%%%%%%%%%%%%%%%%%
\begin{figure}[t!]
\centering
\includegraphics[width=\columnwidth]{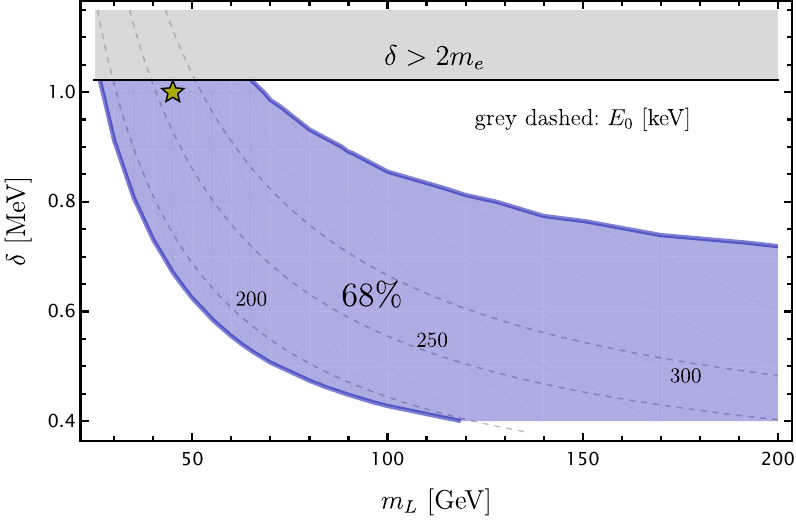}
\caption{The $68\%$ region in the $(m_L,\delta)$ plane at $m_{A'} = 1\GeV$, with the couplings profiled out: blue fill is $-2\Delta\ln\mathcal{L} < 2.30$.  Grey dashed curves are contours of the peak energy $E_0 = \delta\,m_L/(m_L+m_{\rm Xe})$. The star shows the benchmark point.}
\label{fig:like}
\end{figure}
%%%%%%%%%%%%%%%%%%%%%%%%%%%%%%

The preferred window has splittings close to $2m_e$ and a few tens of GeV mass as seen in Fig.~\ref{fig:like}. The maximum sits at $m_L = 45\GeV$ and $\delta = 1.022\MeV$, with $\hat{n} = 0.85$ events in the LZ window. For $m_L$ below about $50\GeV$, the fit always improves with increasing $\delta$, but going above the electron threshold may require additional model building to be cosmologically stable. From this scan, we adopt the following benchmark
\begin{align}
m_L = 45\GeV, \quad \delta = 1.0\MeV, \quad m_{A'} = 1\GeV \, .
\label{eq:bench}
\end{align}
For this benchmark, the expected number of events in the LZ window is $\hat n = 1.04$.

%%%%%%%%%%%%%%%%%%%%%%%%%%%%
\section{Dark Matter Cosmology}
%%%%%%%%%%%%%%%%%%%%%%%%%%%%%

The LZ measurement does not fully select the parameter space, and thus there is still some freedom in the production mechanism. This dark matter candidate could be populated from secluded freeze-out, fixing the dark gauge coupling, and if no other decay mode exists, also setting $f_H$. Another possibility is freeze-in through the same dark matter portal, in which case the relic abundance also involves the visible-sector coupling that controls the direct detection rate. In this section, we overview these two scenarios.

%%%%%%%%%%%%%%%%%%%%%%%%%%%%%
\subsection{Freeze-out}
%%%%%%%%%%%%%%%%%%%%%%%%%%%%%

The small kinetic mixing forces the annihilation of the two Majorana states to be secluded~\cite{Pospelov:2007mp}: $\chi\chi \to A'A'$, with $\langle\sigma v\rangle = \pi\alpha_D^2/m_L^2$. The current dark matter abundance is then fixed by $g_D$, but the kinetic mixing parameter is unconstrained. At  $T \gg \delta$, the two states are equally populated. Then the exothermic conversion $\chi_H\chi_H\to\chi_L\chi_L$ keeps depleting the excited state after the freeze-out~\cite{Finkbeiner:2009mi,Blennow:2016gde,Bhattiprolu:2023akk}. Its cross section is velocity-independent
\begin{align}
\langle\sigma v\rangle_{H\to L} \simeq \frac{g_D^4\, m_L\, \sqrt{2m_L\delta}}{4\pi\left(2m_L\delta+m_{A'}^2\right)^2} ,
\label{eq:sconv}
\end{align}
and is five orders of magnitude larger than the annihilation cross section, since the mediator is light. We solve
\begin{align}
\frac{d f_H}{dt} = -\langle\sigma v\rangle_{H\to L}\, n_\chi \left[f_H^2 - (1-f_H)^2 e^{-2\delta/T_\chi}\right] ,
\label{eq:boltz}
\end{align}
with $T_\chi = T$ above kinetic decoupling and $T_\chi = T^2/T_{\rm kd}$ below it. The dark matter decouples from the dark photon bath at $T_{\rm kd} \approx 51\MeV$ for the benchmark parameter value. We estimate the depletion of the excited state for the benchmark point to be 
\begin{align}
f_H \approx 6.8\times 10^{-3} \quad (m_{A'} = 1\GeV) ,
\end{align}
and $f_H$ ranges from $f_H \approx 2.7\times 10^{-4}$ at $m_{A'} = 0.1\GeV$ with $T_{\rm kd}\approx 7 \MeV$ to $f_H \approx 0.29$ at $m_{A'} = 5\GeV$ with $T_{\rm kd} \approx 228 \MeV$. For the rate measurement, we only have the freedom of $\epsilon$, and for the benchmark point this fixes to around  $\epsilon = 1.3\times 10^{-6}$. Depletion by $\chi_H e \to \chi_L e$ off the electron bath, which dominates in sub-GeV models~\cite{Finkbeiner:2009mi,Bhattiprolu:2023akk}, is negligible at late times in this parameter region.

Because the relic abundance fixes $\alpha_D$, energy injection around recombination can put part of the parameter space in tension with Planck~\cite{Planck:2018vyg}. The scaling of $\alpha_D$ with $m_L$ pushes the largest injection to lighter masses, leaving $m_\chi$ below roughly $30\GeV$ disfavored~\cite{Slatyer:2015jla}.

%%%%%%%%%%%%%%%%%%%%%%%%%%%%%%%%
\subsection{Freeze-in}
%%%%%%%%%%%%%%%%%%%%%%%%%%%%%%%

In another direction, if the dark sector never thermalizes, the states are populated through $f\bar f \to A'^* \to \chi_L\chi_H$ in equal numbers again~\cite{Hall:2009bx,Heeba:2023bik}. The same combination of couplings links the production and direct detection. Integrating the collision term with the yield dominated at $T \sim m_L$ gives
\begin{align}
\alpha_D\,\epsilon^2 \simeq 1.0\times 10^{-23} \qquad (T_{\rm RH} \gg m_L),
\end{align}
independently of $m_L$. One event at LZ needs $\alpha_D\epsilon^2 = 3.5\times 10^{-17}$ at $f_H = 1/2$, so the coupling required by the event overproduces dark matter by a factor of $3.5\times 10^{6}$.

To avoid overproduction, the reheating temperature must be below the mass, which cuts off production in the Boltzmann tail. For the benchmark of this paper, it requires 
\begin{align}
T_{\rm RH} \simeq \frac{m_L}{9.2} = 4.9 \GeV \, .
\end{align}
Requiring that the dark sector does not equilibrate bounds $\alpha_D \lesssim 2.2\times 10^{-4}$ and pushes the reheating temperature down to $m_L/21$ at that boundary.

%%%%%%%%%%%%%%%%%%%%%%%%%%%%%
\section{Discussion}
%%%%%%%%%%%%%%%%%%%%%%%%%%%%%

%%%%%%%%%%%%%%%%%%%%%%%%%%%%%
\begin{figure}[t!]
\centering
\includegraphics[width=\columnwidth]{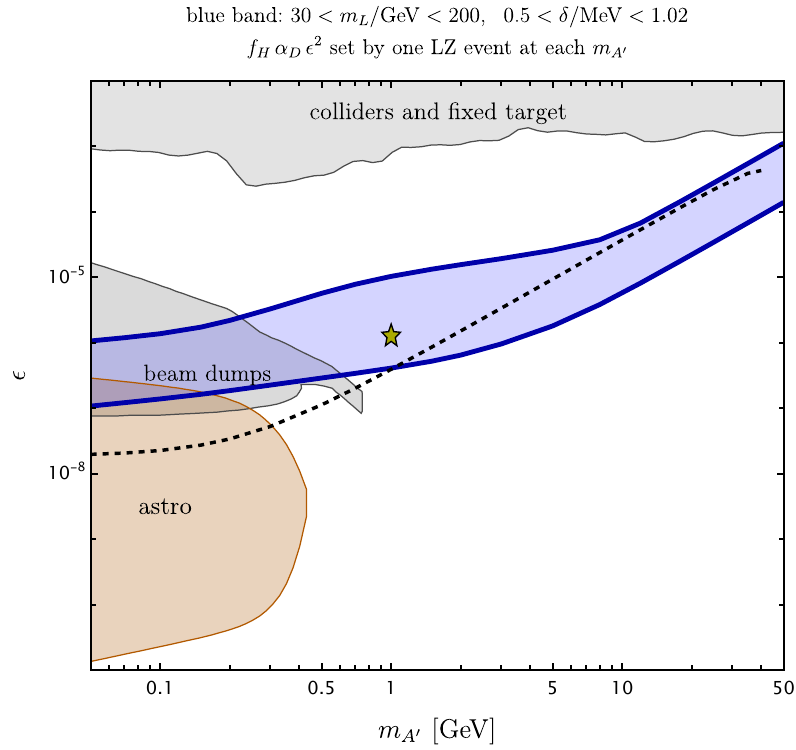}
\caption{Kinetic mixing that reproduces one LZ event as a function of the dark photon mass. The blue band shows the allowed region for freeze-out over the window $30 < m_L/\GeV < 200$ and $0.5 < \delta/\MeV < 1.02$, with $\alpha_D$ fixed by the relic abundance and $f_H$ from \cref{eq:boltz} at each $m_{A'}$. The black dashed curve is freeze-in at the $m_L$ and $\delta$ of \cref{eq:bench}, with $f_H = 1/2$ and $\alpha_D = 2.2\times10^{-4}$. The star shows the benchmark parameter point. The grey regions are excluded by terrestrial searches for a visibly decaying
dark photon~\cite{Ilten:2018crw,Baruch:2022esd}: the upper region is the collider and fixed-target searches~\cite{BaBar:2014zli,NA482:2015wmo,LHCb:2019vmc,KLOE-2:2018kqf,BESIII:2017fwv,Merkel:2014avp,CMS:2019buh}, and the lower wedge the beam dumps~\cite{Bjorken:1988as,Bjorken:2009mm,Davier:1989wz,Blumlein:2011mv,Blumlein:2013cua,Gninenko:2012eq,Tsai:2019buq}. The orange region indicates astrophysical bounds~\cite{Caputo:2025avc}.}
\label{fig:portal}
\end{figure}
%%%%%%%%%%%%%%%%%%%%%%%%%%%%

If we put everything together, we can show the preferred and acceptable window in the $\epsilon$ vs $m_{A'}$ parameter space in Fig.~\ref{fig:portal}. Both freeze-in and freeze-out scenarios are currently allowed as an explanation of this measurement. For dark photons lighter than a GeV, the laboratory and astrophysical constraints reach into the allowed parameter space. For the GeV to 50 GeV window, the parameter space lies in a region that dark photon experiments are not in a position to test; nevertheless, there is a robust prediction from this model in terms of the expected signals at different dark matter experiments with different targets. The signal region depends only on $m_L$ and $\delta$, but the ratio of $E_0$ location for different targets is independent of $\delta$:
\begin{align}
\frac{E_0({\rm Ar})}{E_0({\rm Xe})} = \frac{m_L+m_{\rm Xe}}{m_L+m_{\rm Ar}} = 2.03 ,
\qquad \frac{E_0({\rm Ge})}{E_0({\rm Xe})} = 1.48 .
\end{align}

We can see in Table~\ref{tab:targets} the expected number of events for different targets.  Argon and germanium give roughly three times more events per unit mass than xenon, because xenon is being probed at its diffraction minimum. DEAP-3600~\cite{DEAP:2019yzn}, whose $231$ live-day run corresponds to $0.52$ tonne-years in its $824$ kg fiducial volume, would contain $0.8$ events between $322$ and $931\keV$ of nuclear recoil energy, and DarkSide-20k~\cite{Manthos:2023swh} would collect thirty per twenty tonne-years. Neither collaboration has published an analysis at these energies.  XENONnT with $3.1$ tonne-years and PandaX-4T with $1.54$ tonne-years expect $1.1$ and $0.6$ events at the same energy, if they extend their kinematics\cite{XENON:2025vwd,PandaX:2024qfu}.

%%%%%%%%%%%%%%%%%%%%%%%%%%%
\begin{table}[h!]
\caption{Signal position, kinematic support, and expected events per tonne-year at the benchmark, before detector efficiency. }
\label{tab:targets}
\begin{ruledtabular}
\begin{tabular}{lccc}
Target & $E_0$ [keV] & support [keV] & $N$/(t yr) \\
\hline
W  & 208 & 104-417 & 0.38 \\
Xe & 269 & 138-527 & 0.44 \\
Ge & 398 & 216-733 & 1.38  \\
Ar & 547 & 322-931 & 1.47 \\
\end{tabular}
\end{ruledtabular}
\end{table}
%%%%%%%%%%%%%%%%%%%%%%%%%

%%%%%%%%%%%%%%%%
\section{Conclusion}
%%%%%%%%%%%%%%%%

In this work, we have explored the possibility that the high-energy recoil reported by LZ comes from exothermic dark matter scattering. Within a minimal inelastic dark photon realization, we have identified a preferred region of parameter space with $m_\chi \sim 30-200~\mathrm{GeV}$, and $\delta \sim 0.5-1~\mathrm{MeV}$, where the recoil spectrum is consistent with a single observed event and no additional activity.

The fit favors splittings near the kinematic upper limit, where the spectrum is narrow and peaked. Importantly, the rate depends on the combination $f_H\alpha_D\epsilon^2$, allowing both freeze-out and freeze-in cosmologies to be considered. While standard freeze-in overproduces dark matter for parameters that reproduce the LZ event, a low reheating temperature can restore consistency, whereas secluded freeze-out naturally yields a small but non-negligible excited-state fraction.

In this region of parameter space, the ground state $\chi_L$ is inert, since up-scattering requires $v > \sqrt{2\delta/\mu}$. On xenon, this threshold runs from $2700$ km/s at the light end of our window down to $1100$ km/s in the most favorable corner, $m_\chi = 200\GeV$ with $\delta = 0.5\MeV$, while the fastest particles in the halo reach only roughly $800$ km/s. This also protects against the solar capture bound of Ref.~\cite{Pospelov:2026ewn}, which was derived for the endothermic interpretation and would otherwise apply to $\chi_L$ here as well~\footnote{The $\chi_H$ population can also be captured by down-scattering in the Sun, controlled by the same combination of parameters as the LZ event; however, in the absence of efficient elastic scattering for the $\chi_L$ produced, they keep a large orbit and are not expected to annihilate.}.

This scenario has a distinctive and testable prediction for different target materials. The peak recoil energy scales in a simple, target-dependent way, independent of the splitting for ratios of two targets, leading to higher-energy signals in lighter nuclei such as argon and germanium. As a result, experiments like DEAP-3600 and DarkSide-20k would produce events at energies well above the current xenon window, providing a sharp cross-check of this interpretation. If the observed event is indeed the first glimpse of dark matter, this scenario points toward a metastable dark sector with MeV-scale splittings and GeV-scale mediators, motivating both dedicated analyses of existing data and future searches extending to higher recoil energies.

\acknowledgments
We thank Brian Batell, David McKeen, David Morrissey, Michael Shamma, and Samuel Homiller for useful discussions.

\bibliography{LZDOWN}

\end{document}